\documentclass[reprint,amsmath,amssymb]{revtex4-2}

\usepackage{tikz}
\usepackage{pgfplots}
\usepackage{circuitikz}

\usepackage{hyperref}
\usepackage{graphicx,amsmath,amsfonts}
\usepackage{amssymb}
\usepackage{dcolumn}
\usepackage{mathrsfs}
\usepackage{xcolor}
\usepackage{bm}
\usepackage{enumerate}
\usepackage{pgfplots}
\usepackage{tikz}
\usetikzlibrary{decorations.markings}
\usetikzlibrary{plotmarks}
\usepackage[export]{adjustbox}

\usepackage{braket}
\usepackage{graphicx,amsmath,amsfonts}
\usepackage{dcolumn}
\usepackage{mathrsfs}
\usepackage{xcolor}
\usepackage{calligra}
\usepackage{bm}
\usepackage{enumerate}
\usepackage{pgfplots}
\usepackage{tikz}
\usetikzlibrary{decorations.markings}
\usetikzlibrary{plotmarks}
\usepackage[export]{adjustbox}
\usepackage{stackengine}

\newcommand{\rb}{\mathbf{r}}

\makeatletter
\newsavebox{\@brx}
\newcommand{\llangle}[1][]{\savebox{\@brx}{\(\m@th{#1\langle}\)}%
  \mathopen{\copy\@brx\kern-0.5\wd\@brx\usebox{\@brx}}}
\newcommand{\rrangle}[1][]{\savebox{\@brx}{\(\m@th{#1\rangle}\)}%
  \mathclose{\copy\@brx\kern-0.5\wd\@brx\usebox{\@brx}}}

\colorlet{LightGray}{gray!10}
\makeatother

\definecolor{matlab1}{HTML}{0072BD}
\definecolor{matlab2}{HTML}{D95319}
\definecolor{matlab3}{rgb}{0.4940,0.1840,0.5560}

\newlength{\mywidth}
\newlength{\myheight}
\newlength{\mydelta}
\begin{document}

\title{Quantum emitter interacting with a dispersive dielectric object: \\ a model based on the modified Langevin noise formalism}

\author{Giovanni Miano}
\affiliation{Department of Electrical Engineering and Information Technology, Universit\`{a} degli Studi di Napoli Federico II, via Claudio 21,  Napoli, 80125, Italy}

\author{Loris Maria Cangemi}
\affiliation{Department of Electrical Engineering and Information Technology, Universit\`{a} degli Studi di Napoli Federico II, via Claudio 21,  Napoli, 80125, Italy}

\author{Carlo Forestiere}
\email[]{carlo.forestiere@unina.it}
\affiliation{Department of Electrical Engineering and Information Technology, Universit\`{a} degli Studi di Napoli Federico II, via Claudio 21,  Napoli, 80125, Italy}

\begin{abstract}
In this paper, we model the interaction of a quantum emitter with a finite-size dispersive dielectric object in an unbounded space within the framework of macroscopic quantum electrodynamics, using the modified Langevin noise formalism, without any restrictions on the emitter level structure or
dipole operator. The quantized electromagnetic field consists of two contributions: the medium-assisted field, which accounts for the electromagnetic field generated by the noise polarization currents of the dielectric, and the scattering-assisted field, which takes into account the electromagnetic field incoming from infinity and scattered by the dielectric. We show that the emitter couples to two distinct bosonic baths: a medium-assisted bath and a scattering-assisted bath, each characterized by its own spectral density. We identify the conditions under which the electromagnetic environment composed of these two baths can be effectively replaced by a single bosonic bath, so that the reduced dynamics of the quantum emitter remain unchanged. In particular, when the initial states of the medium- and scattering-assisted baths are thermal states with the same temperature, we find that a single bosonic bath with a spectral density equal to the sum of the medium-assisted and scattering-assisted spectral densities is equivalent to the original electromagnetic environment.
\end{abstract}

\maketitle

\section{Introduction}

The problem of interaction between quantum emitters and arbitrary electromagnetic environments, which are open, dispersive, and absorbing, has drawn significant attention in recent years because of the prospect of altering the physical properties of the emitters (e.g. \cite{chikkaraddy_single-molecule_2016}, \cite{frisk_kockum_ultrastrong_2019}, \cite{gonzalez-tudela_lightmatter_2024},\cite{qin_quantum_2024}). In this scenario, the spectrum of the electromagnetic field is characterized by broad and overlapping resonance peaks embedded into the continuum. 

As losses and dispersion must be considered, quantization of the electromagnetic field constitutes a genuine challenge. Macroscopic quantum electrodynamics has provided a phenomenological recipe for quantizing the electromagnetic field in arbitrary open structures, including dispersive and lossy materials (e.g. \cite{scheel_macroscopic_2008}, \cite{westerberg_perturbative_2023}). It is based on the Langevin noise formalism where, according to the fluctuation-dissipation theorem, it is assumed that the electromagnetic field is produced by the noise polarization current of the dielectric through the dyadic Green function \cite{gruner_green-function_1996}, \cite{dung_three-dimensional_1998}; this electromagnetic field is called the \textit{medium-assisted field} in the literature. Macroscopic quantum electrodynamics is highly versatile and widely used in various research areas such as quantum emitter decay (e.g., \cite{scheel_quantum_1999}, \cite{dung_spontaneous_2000}, \cite{rivera_shrinking_2016}, \cite{hemmerich_influence_2018}, \cite{franke_quantization_2019}, \cite{wang_theory_2020},\cite{feist_macroscopic_2021}), cavity QED (e.g. \cite{dzsotjan_mode-selective_2016}), quantum nanophotonics (e.g. \cite{kurman_tunable_2020}, \cite{feist_macroscopic_2021}), dispersion forces (e.g. \cite{buhmann_casimir-polder_2008}), and fast electron
scattering (e.g. \cite{kfir_optical_2021}). 

Using the Fano method to diagonalize the Hamiltonian, Philbin has provided a microscopic approach to quantizing the electromagnetic field in the presence of a dispersive dielectric object, which fully reproduces and theoretically substantiates the Langevin noise formalism \cite{philbin_canonical_2010}. However, in his approach, the contribution of the vacuum field was disregarded: if one takes the limit of zero electric susceptibility of the dielectric, the free electromagnetic field is not recovered.
Di Stefano \cite{stefano_mode_2001} and Drezet \cite{drezet_quantizing_2017} argued that the original Langevin noise model is incomplete, as it omitted the influence of the fluctuating electromagnetic field coming from infinity and
subsequently scattered by the dielectric object. This observation has triggered renewed interest in the subject \cite{dorier_canonical_2019}, \cite{dorier_critical_2020}, \cite{forestiere_operative_2022}, \cite{forestiere_integral_2023}, \cite{na_numerical_2023}, and a modified Langevin noise formalism explicitly accounting for the fluctuating radiation field incoming from infinity has been analyzed and numerically verified in a specific geometry \cite{na_numerical_2023}.
In particular, starting from the Philbin microscopic model in the Heisenberg picture, Ciattoni \cite{ciattoni_quantum_2024} has recently justified that for an arbitrary dielectric object the quantized electromagnetic field is given by superposition of the medium assisted field and the fluctuating electromagnetic field incoming from infinity and scattered by the dielectric, which we call the \textit{scattering assisted field}. The modified Langevin noise formalism adds the scattering-assisted field to the medium-assisted field in the original Langevin noise model; therefore, it includes both medium- and electromagnetic-field fluctuations on an equal footing. The modified Langevin noise formalism addresses the critiques done in the literature and, therefore, provides a solid foundation for macroscopic quantum electrodynamics.

The analysis of the impact of both the medium-assisted field and the scattering-assisted field on the dynamics of electromagnetic environments that include dispersive dielectrics is crucial to understanding the light-matter interaction mechanism in complex nanophotonics systems. In this paper, we analyze the role of both fields in the interaction of a quantum emitter with a dispersive dielectric. The contributions of the paper are twofold:
(i) We propose a model for a quantum emitter interacting with a dispersive dielectric object based on the modified Langevin noise formalism. We obtain the result that the quantum emitter is coupled with two bosonic baths: a medium-assisted 
bath and a scattering-assisted bath.  Each bath is characterized by a proper continuum spectral density. We used emitter-centered modes to reduce the number of electromagnetic modes of both baths coupled with the emitter \cite{feist_macroscopic_2021}.
(ii) Using this model, we found that the reduced dynamics of the quantum emitter can be described by an equivalent environment with only one bosonic bath, assuming the entire system initially to be in a product state and the initial states for the medium- and scattering-assisted baths to be Gaussian. This equivalence is guaranteed when the expectation values and the two-time correlation
functions of the environment interaction operators of the two environment configurations are equal at all times. In particular, when both baths are initially in thermal states at the same temperature, the spectral density of the equivalent environment is given by the sum of the spectral densities of the matter-assisted bath and of the scattering-assisted bath.

The paper is organized as follows: Section \ref{sec:Model} describes the essence and main features of the modified Langevin noise formalism. Section \ref{sec:BrightDark} presents the emitter-centered mode approach, Section \ref{sec:MinimalRep} introduces and analyzes the model of the quantum emitter interacting with two bosonic baths. Section \ref{sec:Sim} presents one-dimensional numerical simulations of a two-level quantum emitter in a lossy dielectric slab for medium and the assisted baths that are initially in the vacuum
state. A summary and conclusions are given in Section \ref{sec:Conclusion}.

\section{Model}
\label{sec:Model}
A quantum emitter interacts with a dispersive linear dielectric object of arbitrary shape in unbounded space. We denote by $V$ the region occupied by the dielectric, by $\varepsilon_\omega(\mathbf{r})$ its relative permittivity in the frequency domain, and by $\rb_a$ the position vector of the quantum emitter. The combination of the electromagnetic field and the dielectric forms the electromagnetic environment of the emitter.

The Hamiltonian of the entire system, quantum emitter $+$ electromagnetic environment, reads
\begin{equation}
\label{eq:ham1}
    \hat{H} = \hat{H}_{a} + \hat{H}_{em} + \hat{H}_{I},
\end{equation}
where $\hat{H}_{a}$ is the bare emitter Hamiltonian, $\hat{H}_{em}$ is the bare Hamiltonian of the electromagnetic environment, and $\hat{H}_{I}$ is the interaction Hamiltonian. In the multipolar coupling scheme and within the dipole approximation $\hat{H}_{I}$ is given by
\begin{equation}
  \hat{H}_{I}=-\hat{\bold{d}}\cdot\hat{\mathbf{E}}(\rb_a)
\end{equation}
where $\hat{\mathbf{E}}(\rb_a)$ is the electric field operator at the position of the emitter and $\hat{\bold{d}}$ is the electric dipole moment operator of the emitter. We assume that $\hat{\bold{d}}=\hat{{d}}\bold{u}$ where $\bold{u}$ is a stationary unit vector.

In the following, we summarize the modified Langevin noise formalism as formulated in \cite{ciattoni_quantum_2024}.
The electric field operator $\hat{\mathbf{E}}(\rb)$ has two contributions: the medium assisted contribution $\hat{\mathbf{E}}^{(M)}(\rb)$ and the scattering assisted contribution $\hat{\mathbf{E}}^{(S)}(\rb)$,
\begin{equation}
\label{eq:Elect}
    \hat{\mathbf{E}} = \hat{\mathbf{E}}^{(M)} + \hat{\mathbf{E}}^{(S)}.
\end{equation}
The medium-assisted contribution is generated by the noise polarization currents of the dispersive dielectric \cite{gruner_green-function_1996}. The noise polarization current density field is expressed as
\begin{equation}
\label{eq:Jdiel}
   \hat{\mathbf{j}}_{noise}(\rb) =\int_0^\infty d\omega \,\hat{\mathbf{J}}_{\omega} (\rb)+H.c.
\end{equation}
where the monochromatic component $\hat{\mathbf{J}}_{\omega} (\rb)$ is given by
\begin{equation}
\label{eq:Jmon}
  \hat{\mathbf{J}}_{\omega} (\rb) =\sqrt{\frac{\hbar\varepsilon_0 \omega^2}{\pi} \operatorname{Im}\left[\varepsilon_\omega\left(\mathbf{r}\right)\right]}\hat{\mathbf{f}}_{\omega}\left(\mathbf{r}\right),
\end{equation}
$\varepsilon_0$ is the dielectric permittivity in vacuum and $\hat{\mathbf{f}}_{\omega}\left(\mathbf{r}\right)$ is the monochromatic bosonic field operator characterizing the noise of the dielectric, whose support is the region $V$.
Then, the field operator $\hat{\mathbf{E}}^{(M)}(\rb)$ is expressed as
\begin{equation}
\label{eq:Elect0}
   \hat{\mathbf{E}}^{(M)}(\rb) =\int_0^\infty d\omega \, \hat{\mathbf{E}}^{(M)}_{\omega} (\rb)+H.c.,
\end{equation}
where the monochromatic component $\hat{\mathbf{E}}_\omega^{(M)}$ is given by
\begin{equation}
\hat{\mathbf{E}}_\omega^{(M)}(\mathbf{r})= \int_V d^3 \mathbf{r}^{\prime} \, \mathcal{G}_{m \omega}\left(\mathbf{r}, \mathbf{r}^{\prime}\right) \cdot \hat{\mathbf{f}}_{\omega}\left(\mathbf{r}^{\prime}\right),
\end{equation}
and
\begin{equation}
\mathcal{G}_{m\omega}\left(\mathbf{r}, \mathbf{r}^{\prime}\right)=i \frac{\omega^2}{c^2} \sqrt{\frac{\hbar}{\pi \varepsilon_0} \operatorname{Im}\left[\varepsilon_\omega\left(\mathbf{r}^{\prime}\right)\right]} \, \mathcal{G}_\omega\left(\mathbf{r}, \mathbf{r}^{\prime}\right);
\end{equation}
$\mathcal{G}_\omega\left(\mathbf{r}, \mathbf{r}^{\prime}\right)$ is the dyadic Green's function in presence of the dielectric satisfying the
equation
\begin{equation}
\left(\frac{1}{\mu_0}\nabla_\rb \times \nabla_\rb \times-k_\omega^2 \varepsilon_\omega\right) \mathcal{G}_\omega\left(\mathbf{r}, \mathbf{r}^{\prime}\right)=\delta\left(\mathbf{r}-\mathbf{r}^{\prime}\right) I,
\end{equation}
and the boundary condition $\mathcal{G}_\omega\left(\mathbf{r}, \mathbf{r}^{\prime}\right) \rightarrow 0$ for $r, r^{\prime} \rightarrow$ $\infty$, $\mu_0$ is the magnetic permeability in vacuum, $k_\omega=\omega/c$, $c$ is the light velocity in vacuum, and $ I$ is the identity dyad. 

Let be $\mathbf{F}_{\omega \mathbf{n} \nu}(\mathbf{r})$ the solution of equation 
\begin{equation}
\left(\frac{1}{\mu_0} \nabla_\rb \times \nabla_\rb \times-k_\omega^2 \varepsilon_\omega\right) \mathbf{F}_{\omega \mathbf{n} \nu}=0,
\end{equation}
when a plane wave is incoming from infinity
\begin{equation}
\mathbf{F}_{\omega \mathbf{n} \nu}^{(in)}(\rb)=e^{ik_{\omega}\rb\cdot \mathbf{n}}\bold{e}_{\mathbf{n} \nu},
\end{equation}
where $\mathbf{n}$ is the unit vector along the wave vector $\mathbf{k} = k_{\omega}\mathbf{n}$ and $\mathbf{e}_{\mathbf{n}1}$, $\mathbf{e}_{\mathbf{n}2}$ are two mutually orthogonal polarization unit vectors that are orthogonal to $\mathbf{n}$. We introduce the electric field $\mathbf{E}_{\omega \mathbf{n} \nu}(\mathbf{r})$ 
\begin{equation}
\mathbf{E}_{\omega \mathbf{n} \nu}(\mathbf{r}) =\sqrt{\frac{\hbar \mu_0 \omega^3}{16 \pi^3 c}} \mathbf{F}_{\omega \mathbf{n} \nu}(\mathbf{r}).
\end{equation}
The fundamental integral identity \cite{ciattoni_quantum_2024}
\begin{multline}
\label{eq:sum}
\int d^3 \mathbf{r''} \, \mathcal{G}_{m\,\omega}(\mathbf{r}, \mathbf{r''}) \cdot \mathcal{G}_{m \, \omega}^{* T}\left(\mathbf{r}^{\prime}, \mathbf{r''}\right)+ \\ \oint d o_{\mathbf{n}} \sum_\nu \mathbf{E}_{\omega \mathbf{n} \nu}(\mathbf{r}) \mathbf{E}_{\omega \mathbf{n} \nu}^*\left(\mathbf{r}^{\prime}\right)= \frac{\hbar \mu_0 \omega^2}{\pi} \operatorname{Im}\left[\mathcal{G}_\omega\left(\mathbf{r}, \mathbf{r}^{\prime}\right)\right]
\end{multline}
holds, where $o_{\mathbf{n}}=(\theta_{\mathbf{n}},\phi_{\mathbf{n}})$ are the polar angles of the unit vector ${\mathbf{n}}$, $do_{\mathbf{n}}=\sin\theta_{\mathbf{n}} d\theta_{\mathbf{n}} d\phi_{\mathbf{n}}$ is the solid angle differential, the integration is performed over the whole solid angle with $\theta \in [0, \pi]$ and $\phi \in [0, 2 \pi]$. This relation is very important, as we shall see later.

The scattering-assisted contribution $\hat{\mathbf{E}}^{(S)}$ is the fluctuating electromagnetic field incoming from infinity and scattered by the dielectric object. It can be represented through the scattering modes $\mathbf{E}_{\omega \mathbf{n} \nu}(\mathbf{r})$.
Then, $\hat{\mathbf{E}}^{(S)}$ is expressed as
\begin{equation}
\label{eq:Elect1}
   \hat{\mathbf{E}}^{(S)}(\rb) =\int_0^\infty d\omega \,\hat{\mathbf{E}}^{(S)}_{\omega} (\rb)+H.c.,
\end{equation}
where the monochromatic component $\hat{\mathbf{E}}_\omega^{(S)}(\mathbf{r})$ is given by
\begin{equation}
\hat{\mathbf{E}}^{(S)}_\omega(\mathbf{r})=
\oint d o_{\mathbf{n}} \sum_\nu \mathbf{E}_{\omega \mathbf{n} \nu}(\mathbf{r}) \hat{g}_{\omega \mathbf{n} \nu},
\end{equation}
and $\hat{g}_{\omega \mathbf{n} \nu}$ is the monochromatic bosonic operator describing the fluctuation of the radiation incoming from infinity.

%

The bosonic field operators $\hat{\mathbf{f}}_{\omega \lambda}(\mathbf{r})$ and $\hat{g}_{\omega \mathbf{n} \nu}$ are independent. Any possible commutation relations between them vanishes except the fundamental ones
\begin{equation}
{\left[\hat{\mathbf{f}}_{\omega \lambda}(\mathbf{r}), \hat{\mathbf{f}}_{\omega^{\prime} \lambda^{\prime}}^{\dagger}\left(\mathbf{r}^{\prime}\right)\right] } =\delta\left(\omega-\omega^{\prime}\right) \delta_{\lambda \lambda^{\prime}} \delta\left(\mathbf{r}-\mathbf{r}^{\prime}\right) I,
\end{equation}
\begin{equation}
{\left[\hat{g}_{\omega \mathbf{n} \nu}, \hat{g}_{\omega^{\prime} \mathbf{n}^{\prime} \nu^{\prime}}^{\dagger}\right] } =\delta\left(\omega-\omega^{\prime}\right) \delta\left(o_{\mathbf{n}}-o_{\mathbf{n}^{\prime}}\right) \delta_{\nu \nu^{\prime}},
\end{equation}
where $\delta\left(o_{\mathbf{n}}-o_{\mathbf{n}^{\prime}}\right)=\delta\left(\theta_{\mathbf{n}}-\theta_{\mathbf{n}}^{\prime}\right) \delta\left(\varphi_{\mathbf{n}}-\varphi_{\mathbf{n}}^{\prime}\right) / \sin \theta_{\mathbf{n}}$. These commutation relations guarantee the canonical commutation relations for the electromagnetic field and the continuum of harmonic oscillator fields of the Philbin microscopic model on which the modified Langevin noise formalism is based \cite{ciattoni_quantum_2024}. In particular, the monochromatic component of the electric field operator $\hat{\mathbf{E}}_\omega(\mathbf{r})=\hat{\mathbf{E}}_\omega^{(M)}(\mathbf{r})+\hat{\mathbf{E}}_\omega^{(S)}(\mathbf{r})$ satisfies the commutation relation
\begin{equation}
{\left[\hat{\mathbf{E}}_\omega(\mathbf{r}), \hat{\mathbf{E}}_{\omega^{\prime}}^{\dagger}\left(\mathbf{r}^{\prime}\right)\right]}=\frac{\hbar \mu_0 \omega^2}{\pi} \operatorname{Im}\left[\mathcal{G}_\omega\left(\mathbf{r}, \mathbf{r}^{\prime}\right)\right]\delta\left(\omega-\omega^{\prime}\right).
\end{equation}

The bare electromagnetic environment Hamiltonian is given by \cite{ciattoni_quantum_2024}
\begin{multline}
\hat{H}_{em}=\int_0^{\infty} d \omega \hbar \omega\left[\int_V d^3 \, \mathbf{r} \,\hat{\mathbf{f}}_{\omega}^{\dagger}(\rb) \cdot \hat{\mathbf{f}}_{\omega}(\rb)+ \right. \\ \left. \oint d o_{\mathbf{n}} \sum_\nu \hat{g}_{\omega \mathbf{n} \nu}^{\dagger} \hat{g}_{\omega \mathbf{n} \nu}\right].
\end{multline}
The operators $\hat{\mathbf{f}}_{\omega}^{\dagger}$, $\hat{\mathbf{f}}_{\omega}$ and $\hat{g}_{\omega \mathbf{n} \nu}^{\dagger}$, $\hat{g}_{\omega \mathbf{n} \nu}$ can be viewed as creation and annihilation operators of two different kinds of polaritons, medium- and scattering-assisted polaritons.

The expression of the electric field \eqref{eq:Elect} differs from that
considered in the Langevin noise formalism (e.g., \cite{westerberg_perturbative_2023}, \cite{feist_macroscopic_2021}) due to the scattering-assisted field contribution. The fundamental integral relation \eqref{eq:sum} differs from that considered in the Langevin noise formalism due to the second term on the left-hand side: it is a surface term that contains the asymptotic
amplitude of the dyadic Green function expressed through the vector field $\mathbf{E}_{\omega \mathbf{n} \nu}(\mathbf{r})$ \cite{ciattoni_quantum_2024}.
The inclusion of the scattering-assisted field and the correct evaluation of the integral $\int d^3 \mathbf{r''} \, \mathcal{G}_{m\,\omega}(\mathbf{r}, \mathbf{r''}) \cdot \mathcal{G}_{m \, \omega}^{* T}\left(\mathbf{r}^{\prime}, \mathbf{r''}\right)$, which also takes into account the contribution of the surface term (e.g., \cite{na_numerical_2023}),
 have addressed the critique of the Langevin noise formalism raised by Refs. \cite{drezet_equivalence_2017,dorier_canonical_2019,dorier_critical_2020}. The expression of the bare electromagnetic environment Hamiltonian also differs from that considered in the Langevin noise formalism because there are two bosonic baths. In the limit of non-dispersive dielectric, the modified Langevin noise formalism reduces to the quantum optic model introduced by Glauber and Lewenstein \cite{glauber_quantum_1991}.

\section{Bright and dark modes}
\label{sec:BrightDark}

We now introduce linear transformations of the bosonic field operators $\hat{\mathbf{f}}_{\omega}$ and $\hat{g}_{\omega \mathbf{n} \nu}$ such that in the new basis, only a minimal
number of bosonic oscillators couples with the emitter \cite{feist_macroscopic_2021}.

We start with the representation of $\hat{\mathbf{f}}_{\omega}(\rb)$. We consider the monochromatic operator $\hat{A}_{\omega}$ defined as
\begin{equation}
\hat{A}_{\omega}=\int_{V}d^3\mathbf{r}  \,  \boldsymbol{\alpha}_{\omega}(\rb)\cdot \hat{\mathbf{f}}_{\omega}(\mathbf{r})
\end{equation}
where
\begin{equation}
\boldsymbol{\alpha}_{\omega}(\rb)=\frac{\bold{u} \cdot \mathcal{G}_{m \, \omega }(\mathbf{r}_a, \mathbf{r})}{g_M(\omega)}
\end{equation}
and $g_M(\omega)$ is an arbitrary normalization parameter. We choose $g_M(\omega)$ in such a way that the commutator between $\hat{A}_{\omega}$ and $\hat{A}_{\omega}^\dagger$ is
\begin{equation}
{\left[\hat{A}_{\omega}, \hat{A}_{\omega^{\prime}}^{\dagger}\right] } =\delta\left(\omega-\omega^{\prime}\right),
\end{equation}
and obtain
\begin{equation}
g_M(\omega)=\sqrt{ \int_V{d^3 \mathbf{r} \, \bold{u}\cdot [\mathcal{G}_{m \,\omega}(\rb_a,\rb}) \cdot \mathcal{G}_{m \, \omega}^{* T}(\rb_a,\rb) ] \cdot \bold{u}} \,.
\end{equation}
Then, the contribution of the medium assisted field to $\hat{H}_{I}$ is expressed as
\begin{equation}
\hat{H}_{I}^{(M)}=-\hat{d} \left[ \int_0^\infty d\omega g_M(\omega) \hat{A}_\omega +H.c. \right].
\end{equation}
On the other hand, we can always express the field operator $\hat{\mathbf{f}}_{\omega}(\mathbf{r})$ as
\begin{equation}
\hat{\mathbf{f}}_{\omega}(\mathbf{r})=\boldsymbol{\alpha}_{\omega}^*(\rb) \hat{A}_\omega + \sum_m [\boldsymbol{\alpha}_{\omega}^m(\rb)]^*\hat{A}_{\omega}^m,
\end{equation}
where the orthonormal set of vector fields $\{\boldsymbol{\alpha}_{\omega}^m(\rb)\}$ span the subspace orthogonal to $\boldsymbol{\alpha}_{\omega}(\rb)$, that is, $ \int_Vd\rb^3 [\boldsymbol{\alpha}_{\omega}^{m}(\rb)]^* \cdot \boldsymbol{\alpha}_{\omega}(\rb) = 0 $. Note that each $\boldsymbol{\alpha}_{\omega}^m(\rb)$ does not couple to the emitter; $\hat{A}_\omega$ is the emitter-centered bright mode of the medium-assisted field, while $\{\hat{A}_{\omega}^m\}$ are an infinite number of dark modes. Then, the contribution of the medium assisted electromagnetic field to $\hat{H}_{em}$ is given by
\begin{equation}
\hat{H}_{em}^{(M)}= \int_0^\infty d\omega \hbar \omega \hat{A}_\omega ^\dagger \hat{A}_\omega +  \int_0^\infty d\omega \hbar \omega \sum_m (\hat{A}_{\omega}^{m})^\dagger \hat{A}_{\omega}^m.
\end{equation}

We now consider the representation of $\hat{g}_{\omega \mathbf{n} \nu}$. We introduce the monochromatic operator $\hat{B}_{\omega}$ defined by
\begin{equation}
\hat{B}_{\omega}=\oint d o_{\mathbf{n}} \sum _\nu \, {\beta}_{\omega \mathbf{n} \nu } \hat{g}_{\omega \mathbf{n} \nu },
\end{equation}
where
\begin{equation}
{\beta}_{\omega \mathbf{n} \nu } =\frac{\bold{u} \cdot \mathbf{E}_{\omega \mathbf{n} \nu}(\mathbf{r}_a)}{g_S(\omega)}.
\end{equation}
Here, $g_S(\omega)$ is an arbitrary normalization real parameter chosen such that the commutator relation
\begin{equation}
{\left[\hat{B}_{\omega}, \hat{B}_{\omega^{\prime}}^{\dagger}\right] } =\delta\left(\omega-\omega^{\prime}\right)
\end{equation}
holds. Thus, we obtain for $g_S(\omega)$
\begin{equation}
g_S(\omega)=\sqrt{ \oint do_{\mathbf{n}} \, \mathbf{u} \cdot [\sum_\nu \mathbf{E}^*_{\omega \mathbf{n} \nu}(\mathbf{r}_a) \mathbf{E}_{\omega \mathbf{n} \nu}(\mathbf{r}_a)] \cdot \bold{u}} \,.
\end{equation}
Then, the contribution of the scattered assisted field to $\hat{H}_{I}$ is given by
\begin{equation}
\hat{H}_{I}^{(S)}=-\hat{d} \left[ \int_0^\infty d\omega g_S(\omega)\hat{B}_\omega +H.c.\right].
\end{equation}
On the other hand, the field operator $\hat{g}_{\omega \mathbf{n} \nu}$ can always be expressed as
\begin{equation}
\hat{g}_{\omega \mathbf{n} \nu}= {\beta}_{\omega \mathbf{n} \nu }^* \hat{B}_\omega + \sum_m [{\beta}_{\omega \mathbf{n} \nu}^m]^* \hat{B}_{\omega}^m,
\end{equation}
where  $\{{\beta}_{\omega \mathbf{n} \nu}^m\}$ is 
an orthonormal set of vector fields spanning the subspace orthogonal to ${\beta}_{\omega \mathbf{n} \nu }$, that is, $\int do_{\mathbf{n}} \sum_\nu [{\beta}_{\omega \mathbf{n} \nu}^m]^* {\beta}_{\omega \mathbf{n} \nu} = 0 $. Note that every ${\beta}_{\omega \mathbf{n} \nu}^m$ does not couple to the emitter; $\hat{B}_\omega$ is the emitter-centered bright mode of the scattering assisted field, and $\{\hat{B}_{\omega}^m\}$ are an infinite number of dark modes. Consequently, the contribution of the scattering assisted field to $\hat{H}_{em}$ is expressed as
\begin{equation}
\hat{H}_{em}^{(S)}= \int_0^\infty d\omega \hbar \omega \hat{B}_\omega ^\dagger \hat{B}_\omega +  \int_0^\infty d\omega \hbar \omega \sum_m (\hat{B}_{\omega}^m)^\dagger \hat{B}_{\omega}^m.
\end{equation}

Using the above results, the Hamiltonian of the entire system reads
\begin{equation}
\hat{H} = \hat{H}_{a} + \hat{H}_{E} + \hat{H}_{I} + \hat{H}^{(dark)}
\end{equation}
where
\begin{equation}
\hat{H}_{E} = \int_0^\infty d\omega \hbar \omega ( \hat{A}_\omega ^\dagger \hat{A}_\omega+ \hat{B}_\omega ^\dagger \hat{B}_\omega),
\end{equation}
\begin{equation}
\hat{H}_{I} = \hat{H}_{I}^{(M)}+\hat{H}_{I}^{(S)},
\end{equation}
and
\begin{equation}
\hat{H}^{(dark)}=\int_0^\infty d\omega \hbar \omega \sum_m [(\hat{A}_{\omega}^m)^\dagger \hat{A}_{\omega}^m+(\hat{B}_{\omega}^m)^\dagger \hat{B}_{\omega}^m].
\end{equation}

The real functions $g_M(\omega)$ and $g_S(\omega)$ are not independent, in fact, we have as a consequence of \eqref{eq:sum}
\begin{equation}
\label{eq:gequiv}
g_M^2 (\omega)    +    g_S^2 (\omega) =\frac{\hbar \mu_0 \omega^2}{\pi} \bold{u}\cdot \operatorname{Im}\left[\mathcal{G}_\omega\left(\mathbf{r}_a, \mathbf{r}_a\right)\right]\cdot\bold{u} .
\end{equation}

\section{Reduced Hamiltonian}
\label{sec:MinimalRep}

Since the dark modes are
decoupled from the rest of the system, they do not affect the
dynamics of the emitter and can be dropped, giving the reduced Hamiltonian
\begin{equation}
\hat{H}_{red} = \hat{H}_{a} + \hat{H}_{E} + \hat{H}_{I}.
\label{eq:Hred}
\end{equation}
If the dark modes are
initially excited, including them might be necessary to fully describe the state of the system. 

For our purpose, it is convenient express $\hat{H}_I$ as
\begin{equation}
\hat{H}_{I} = -\hat{d}\hat{F},
\end{equation}
where $\hat{F}$ is the effective electromagnetic environment interaction operator
\begin{equation}
\hat{F} = \hat{F}_M+ \hat{F}_S
\end{equation}
with 
\begin{equation}
\hat{F}_{M} =\int_0^\infty d\omega g_M(\omega) \hat{A}_\omega +H.c.,
\end{equation}
\begin{equation}
\hat{F}_{S} =\int_0^\infty d\omega g_S(\omega) \hat{B}_\omega +H.c. \, .
\end{equation}
$\hat{F}_M$ is the operator through which the medium assisted bath interact with the emitter and $\hat{F}_S$
is the operator through which the scattering assisted bath interacts with the emitter.  We note that the interaction between the emitter and the electromagnetic environment is characterized by two spectral densities. Let us indicate with $m$ the transition dipole moment of the quantum emitter. The medium-assisted spectral density $\mathcal{J}_M(\omega)=[g_M(\omega) \, m/\hbar]^2$ is related to the coupling strength $g_M(\omega)$ of the emitter-centered mode $\hat{A}_{\omega}$. The scattering-assisted spectral density $\mathcal{J}_S(\omega)=[g_S(\omega) \, m/\hbar]^2$ is related to the coupling strength $g_S(\omega)$ of the emitter-centered  mode $\hat{B}_{\omega}$.

The quantum emitter can be described as an open quantum system that interacts with two independent bosonic baths characterized by two different spectral densities. Let us introduce the expectation value $F(t)$ and the two-time correlation function $C(t+\tau;t)$ of the operator $\hat{F}$ as given by the
evolution of the electromagnetic environment with no coupling to the quantum emitter (i.e., electromagnetic environment in free evolution),
\begin{equation}
F(t) = \text{Tr}_E \left[\hat{U}^\dagger_E(t)\hat{F} \hat{U}_E(t) \hat{\rho}_{E}(0) \right],
\end{equation}
\begin{multline}
C(t+\tau;t) =  \\ \text{Tr}_E \left[\hat{U}^\dagger_E(t+\tau)\hat{F} \hat{U}_E(t+\tau) \hat{U}^\dagger_E(t)\hat{F}\hat{U}_E(t)\hat{\rho}_{E}(0) \right].
\end{multline}
where $\hat{U}_E(t)=\exp(-i\hat{H}_{E}t/\hbar)$. For initial product states of the entire system, $\hat{\rho}(0)=\hat{\rho}_a(0) \otimes \hat{\rho}_E(0)$, where $\hat{\rho}_a (0) $ and $\hat{\rho}_E(0)$ are the initial density operators of the emitter and of the environment, respectively, and assuming the environment to initially be in a Gaussian state, the evolution of the reduced density operator of the emitter $\hat{\rho}_a(t)$ depends only on $F(t)$ and $C(t+\tau;t)$ (e.g., \cite{h_p_breuer_and_f_petruccione_theory_2002}). This fundamental property allows the design of an \textit{equivalent environment} with only a single bosonic bath to compute the time evolution of the reduced density operator of the emitter. Let us indicate with $F_{eq}(t)$ and $C_{eq}(t+\tau;t)$ the expectation value and the two-time correlation function of the interaction operator of the equivalent environment considered in free evolution, with $F_M(t)$ and $F_S(t)$ the expectation value of $\hat{F}_M$ and $\hat{F}_S$ and with $C_M(t+\tau;t)$ and $C_S(t+\tau;t)$ the corresponding two-time correlation functions when the electromagnetic environment is in free evolution. Then, we have:
\begin{equation}
F_{eq}(t) = F_M(t) + F_S(t)
\end{equation}
and
\begin{multline}
C_{eq}(t+\tau;t) = C_M(t+\tau;t) + C_S(t+\tau;t)+ \\\left[F_M(t+\tau) F_S(t)+ F_S(t+\tau) F_M(t)\right].
\end{multline}
If the expectation values of the two interacting operators are equal to zero, we have 
\begin{equation}
C_{eq} = C_M + C_S.
\label{eq:SumC}
\end{equation}
Moreover, if both bosonic baths are initially in a thermal state, we obtain
\begin{equation}
C_\alpha(t)= \left(\frac{\hbar}{m} \right)^2\int_0^{\infty} \mathrm{d} \omega {\mathrm{\mathcal{J}}_\alpha(\omega)} \Theta(\omega t; \beta_{\alpha} \hbar\omega)
\end{equation}
where
\begin{equation}
\Theta(\omega t; \beta_{\alpha} \hbar\omega)=\operatorname{coth}\left(\frac{\beta_\alpha \hbar\omega}{2}\right) \cos (\omega \mathrm{t})-{i} \sin (\omega \mathrm{t}),
\end{equation}
and $\alpha = M, S$; $\mathcal{J}_\alpha$ is the spectral density characterizing the coupling of the $\alpha$-type bosonic bath to the emitter, $\beta_\alpha=1/k_B T_\alpha$ and $T_\alpha$ is the temperature of the $\alpha$-type bosonic bath. The Fourier transform of the correlation function $C_\alpha(t)$, i.e., the power spectrum, is given by $S_\alpha(\omega)=\int_{-\infty}^{+\infty} dt e^{i\omega t} C_\alpha(t)=\pi \hbar^2 {\mathrm{\mathcal{J}}_\alpha(\omega)}[1+\coth(\beta_{\alpha}\hbar\omega)]$.

When the temperatures of the two bosonic baths are equal ($T_M=T_S=T_0$), we obtain
\begin{equation}
C_{eq}(t)= \left(\frac{\hbar}{m} \right)^2\int_0^{\infty} \mathrm{d} \omega {\mathrm{\mathcal{J}}_{eq}(\omega)} \Theta(\omega t; \beta_{0}\hbar \omega)
\end{equation}
where $\beta_0=1/k_BT_0$ and 
\begin{equation}
\label{eq:equiv}
\mathcal{J}_{eq}(\omega)=\mathcal{J}_M(\omega)+\mathcal{J}_S(\omega)
\end{equation}
 is the spectral density of the equivalent single bath. Using \eqref{eq:gequiv}, we obtain
\begin{equation}
\mathcal{J}_{eq}(\omega)=\frac{m^2 \mu_0 \omega^2}{\pi\hbar} \bold{u} \cdot \operatorname{Im}\left[\mathcal{G}_\omega\left(\mathbf{r}_a, \mathbf{r}_a\right)\right]\cdot \bold{u}.
\label{eq:OneSD}
\end{equation}
In the regime of weak coupling, i.e., when the electromagnetic environment can be approximated as a
Markovian bath, the spontaneous emission rate at an
emitter frequency $\omega_a$ is given by $2\pi \mathcal{J}_{eq}(\omega_a)$.
It is crucial to note that, in the literature on the interaction between quantum emitters and dispersive dielectric objects based on the Langevin noise formalism, which omits the scattering-assisted field, the reduced dynamics of the quantum emitter is studied by using the spectral density $\mathcal{J}_{eq}(\omega)$ given by \eqref{eq:OneSD}. How is it possible that two different models give the same result for the expression of the spectral density?  This is due to the fact that, although the scattering-assisted bath is ignored in the Langevin noise formalism, a surface term is omitted in the calculation of the integral $\int d^3 \mathbf{r''} \, \mathcal{G}_{m\,\omega}(\mathbf{r}, \mathbf{r''}) \cdot \mathcal{G}_{m \, \omega}^{* T}\left(\mathbf{r}^{\prime} \mathbf{r''}\right)$ (see the second term in the l.h.s. of Eq. \eqref{eq:sum}), and this leads to the wrong relation $\int d^3 \mathbf{r''} \, \mathcal{G}_{m\,\omega}(\mathbf{r}, \mathbf{r''}) \cdot \mathcal{G}_{m \, \omega}^{* T}\left(\mathbf{r}^{\prime}, \mathbf{r''}\right)=\frac{\hbar\omega^2}{\pi} \operatorname{Im}\left[\mathcal{G}_\omega\left(\mathbf{r}, \mathbf{r}^{\prime}\right)\right]$. 
This result clarifies a much-debated issue in the literature.

When the baths are in non-equilibrium states, such as when the temperatures of the two baths are different, the Langevin noise formalism and the modified Langevin noise formalism yield different reduced dynamics for the quantum emitter.

\begin{figure*}[t]
\includegraphics[width=\textwidth]{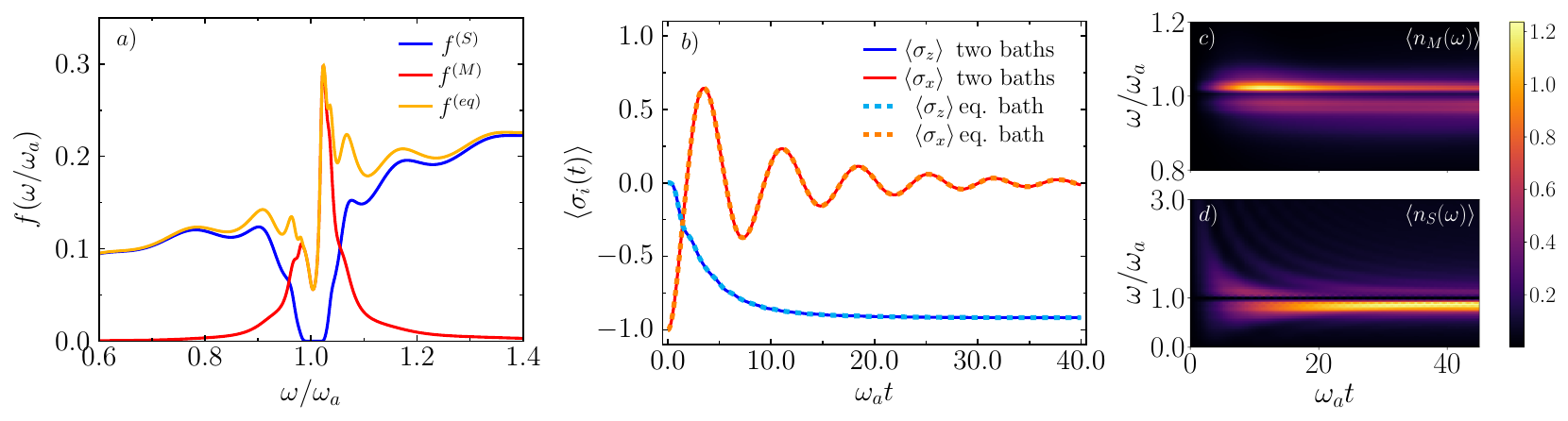}
\caption{(a) Normalized spectral density of the scattering (S), medium (M) and equivalent (eq) baths plotted against $\omega/\omega_a$. (b) Expectation values of $\hat{\sigma}_x$ and $\hat{\sigma}_z$ plotted versus time. Case i) Solid lines: the emitter couples to the medium and scattering baths, prepared at $t=0$ in their vacuum states. Case ii) Dashed lines: the emitter couples to a single equivalent bath with spectral density $\mathcal{J}_{eq}=\mathcal{J}_{{S}} +\mathcal{J}_{{M}}$, which at $t=0$ is in its vacuum state. Expectation values of the occupation numbers of the medium bath modes $\hat{n}^{{M}}_{\omega}$ (c) and of the scattering bath modes $\hat{n}^{{S}}_{\omega}$ (d), plotted versus mode frequency and time. The parameters are the same as in (b).}
\label{fig:scomp}
\end{figure*}

\section{Simulation results}
\label{sec:Sim}

We now present some results of the simulation of the evolution of a two-level quantum emitter located at the center of a homogeneous dielectric slab, obtained by applying the modified Langevin noise formalism. To verify the equivalence condition \eqref{eq:equiv}, we assume that the medium and the scattering baths are initially in their respective vacuum states, while the emitter is initially in a pure state. The dielectric slab has thickness $\ell$ and electric susceptibility $\chi(\omega) = (\omega_p/\omega_0)^2/[{1 - (\omega/\omega_0)^2 - i (\omega /\omega_0) (\gamma /\omega_0)}]$. As in ref. \cite{na_numerical_2023}, we chose \(\omega_p/\omega_0 = 0.2\), \(\gamma/\omega_0 = 0.01\) and $(\omega_0/c)\ell=31.25$. We use $\hat{\sigma}_i$,
with $i=x, y, z$, to denote the Pauli matrices,
and ${\ket{\pm}}$ to denote the eigenstates of $\hat{\sigma}_z$, that is, $\hat{\sigma}_{z}\ket{\pm}=\pm\ket{\pm}$. The bare Hamiltonian of the two-level quantum emitter reads $\hat{H}_{a}={\hbar\omega_a}(\hat{\sigma}_{z}/2)$ where $\omega_a$ is the bare transition frequency. The electric dipole moment operator is given by $\hat{d} = m \hat{\sigma}_{x}$.

The medium and scattering baths are initially prepared in thermal equilibrium states at zero temperature. The emitter is initially prepared in the pure state $\hat{\rho}_{a}(0)=\ket{x}\bra{x}$ where $\ket{x}=(1/\sqrt{2}) (\ket{+}-\ket{-})$ is an eigenstate of $\hat{\sigma}_x$. Since the initial state of the entire system does not coincide with an eigenstate of $\hat{H}_{red}$, given by \eqref{eq:Hred}, the entire system evolves for $t>0$ into a correlated state of the emitter and both baths \cite{perarnau-llobet_strong_2018,del_pino_tensor_2018}. When the temperatures of the two baths are identical, the reduced dynamics of the emitter can also be evaluated using an equivalent single bath with spectral density $\mathcal{J}_{eq}(\omega)$. Nevertheless, it is fair to stress that the original model of the electromagnetic environment with two baths allows the direct evaluation of the statistics of its physical variables.
To show these features, we simulated the unitary dynamics of the state $| \psi (t) \rangle $ of the whole system employing the Matrix Product States technique \cite{schollwock_density-matrix_2011,prior_efficient_2010,weimer_simulation_2021,fishman_itensor_2022} from which the density operator  $\hat{\rho}(t) = | \psi  (t) \rangle \langle \psi (t) |$  is immediately obtained. 

We used a one-dimensional model for the quantum emitter and the dielectric slab to calculate the medium and scattered assisted electric fields \cite{na_numerical_2023}. In Fig. \ref{fig:scomp}(a), we show the frequency behavior of the spectral densities $\mathcal{J}_{S}(\omega),\mathcal{J}_{M}(\omega)$ and $\mathcal{J}_{eq}(\omega)$ expressed as $\mathcal{J}_{\alpha}(\omega)=\eta \, \omega_a f^{(\alpha)}({\omega}/{\omega_a})$ with $\alpha=S,M, eq$, where $\eta =  {\zeta_0 \, m^2}/({\Sigma \hbar})$, $\zeta_0=\sqrt{\mu_0/\varepsilon_0}$ and $\Sigma$ is an effective area. We choose the bare emitter transition frequency $\omega_a$ equal to the resonance frequency of the dielectric slab $\omega_{0}$, $\omega_a=\omega_0$. Although $\mathcal{J}_{{M}}(\omega)$ shows a doubly-peaked structure in a narrow frequency interval centered at $\omega_{a}$, $\mathcal{J}_{{S}}(\omega)$ extends throughout the frequency spectrum. In the one-dimensional model, $\mathcal{J}_{{S}}(\omega)$ is approximately zero around $\omega_a$ because the scattering-assisted field is almost completely reflected by the slab at the resonance frequency of the dielectric. Far from the resonance frequency, $\mathcal{J}_{{S}}(\omega)$ increases linearly with frequency because the plane waves that come from infinity completely penetrate the dielectric slab.

We performed simulations of the evolution of $\hat{\rho}(t)$ considering the instances of an emitter coupled with: Case i) two different baths, each described by $\mathcal{J}_{S}(\omega)$ and $\mathcal{J}_{M}(\omega)$; Case ii) a single equivalent bath with $\mathcal{J}_{eq}(\omega)$. In both cases, we assumed $\eta= 2 \pi \times 0.05$. We used the Matrix Product States technique,  applying a cut-off frequency $\omega_c= 4\omega_{0}$ and using $N=500$ discrete bosonic modes for each bath, with a maximum local dimension of $n_{{max}}=3$.
In Fig. \ref{fig:scomp}(b), we plot the expectation values $\langle \hat{\sigma}_{x}\rangle (t)=\text{Tr}[\hat{\sigma}_{x}\hat{\rho}(t)]$ and $\langle \hat{\sigma}_{z}\rangle (t)=\text{Tr}[\hat{\sigma}_{z}\hat{\rho}(t)]$ versus time. The evolution of $\langle \hat{\sigma}_{y}\rangle (t)$, not shown here, differs from that of $\langle \hat{\sigma}_{x}\rangle (t)$ for a phase shift of roughly $\pi/2$. As expected, for the chosen initial states of the bath, the dynamics of the observables coincide in the two cases, indicating that the influence of the dielectric slab on the reduced dynamics of the emitter can be effectively simulated with a single, equivalent bath. The emitter dynamics show that the population of the $\ket{-}$ eigenstate increases at the expense of the population of the $\ket{+}$ eigenstate. However, the reduced state $\rho_{a}(t)$ does not converge to the ground state of the emitter for long times. Indeed, this behavior can be attributed to the quantum correlations established between the emitter and the baths. At the same time, the coherence of the emitter state decreases over time, and the purity of the reduced state at the final times depends on the coupling strength. 

In Figs. \ref{fig:scomp}(c) and \ref{fig:scomp}(d), we plot the time evolution of the expectation values of the number operators for the medium and scattering bath modes at frequency $\omega$, $\langle{n^{M}_{\omega}\rangle(t)}=\text{Tr}[A^{\dagger}_{\omega} A_{\omega}\hat{\rho}(t)]$ and $\langle{n^{S}_{\omega}\rangle(t)}=\text{Tr}[B^{\dagger}_{\omega} B_{\omega}\hat{\rho}(t)]$. Once the dynamics start from the product state, the bath modes start to get increasingly populated. The scattering bath modes show significant population increases at low frequencies, after a transient time of the order of $10/\omega_{a}$, they reach a steady state, with a maximum below $\omega_{a}$, which is followed by a dark window around $\omega_{a}$ due to the resonance of the dielectric slab. In contrast, the medium bath modes show a non-trivial time evolution of $\langle{n^{M}_{\omega}\rangle(t)}$: during the transient dynamics the modes with $\omega\geq \omega_{a}$ increase their populations before converging towards their stationary values, where the relative weights of the mode populations with frequencies $\omega\geq \omega_{a}$ are changed.

\section{Conclusions and outlook}
\label{sec:Conclusion}
In summary, we have proposed a model for a quantum emitter that interacts with a finite-size dispersive dielectric object in an unbounded space based on the modified Langevin noise formalism, without restrictions on the emitter level structure or
dipole operators. The electromagnetic environment is composed of two bosonic baths: the medium-assisted bath and the scattering-assisted bath. The medium-assisted bath describes the electromagnetic field generated by the noise polarization currents of the dielectric; the scattering-assisted bath describes the radiation incoming from infinity and scattered by the dielectric. We used emitter-centered modes to reduce the number of electromagnetic modes of both baths coupled to the emitter. Each bath is characterized by a proper continuum spectral density. The model for the Hamiltonian proposed in \eqref{eq:Hred} allows us to treat the evolution of the reduced dynamics of the emitter for arbitrary electromagnetic environments, including dispersive dielectric objects, and for arbitrary initial quantum states of the two bosonic baths, for instance, initial states with non-zero expectation value, or thermal states with different temperatures.

For an initial product state and an initial Gaussian state for the electromagnetic environment, the two-bath electromagnetic environment can be replaced by an effective single bosonic bath. The interaction operator of the effective single bath is prescribed to have the same expectation value and the same two-time correlation function as the interaction operator of the original environment.
When the two baths are in a thermal state with the same temperature, the effective single bath can be characterized by a spectral density equal to the sum of the medium-assisted spectral density and the scattering-assisted spectral density. It is related to the Green function through the relation $\mathcal{J}(\omega)=\frac{m^2 \mu_0 \omega^2}{\pi\hbar} \bold{u}\cdot \operatorname{Im}\left[\mathcal{G}_\omega\left(\mathbf{r}_a, \mathbf{r}_a\right)\right]\cdot\bold{u}$. In the literature based on the Langevin noise formalism, this expression is widely used; however, the conditions under which it remains valid are not always clearly stated. When the baths are in non-equilibrium states, e.g., when the temperatures of two baths are different, it is not possible to introduce an equivalent spectral density, and a description in terms of an equivalent single bath has to rely on Eq. \eqref{eq:SumC}. These conclusions suggest that new physics can be found from the investigation of the dynamic of a quantum emitter in the presence of two baths in non-equilibrium states, for which the equivalent single-bath spectral density can no longer be defined.

We envision that, similarly to what we have shown in this paper, the investigation of models that incorporate both medium-assisted and scattering-assisted baths will have significant implications for various research fields, including cavity QED, quantum nanophotonics, dispersion forces, and fast electron scattering.



\begin{acknowledgements}
This work was supported by Ministero dell’Università
e della Ricerca under the PNRR Projects No. CN00000013-ICSC.
\end{acknowledgements}

\end{document}